\documentclass[preprint,amssymb,amsmath,amsfonts,superscriptaddress,12pt]{revtex4-1}
\usepackage[utf8]{inputenc}
\usepackage{mathtools}
\usepackage{graphicx}
\usepackage{caption}
\usepackage{hyperref}
\usepackage{amssymb}
\usepackage{color} 
\usepackage{soul} 

\makeatletter
\let\saved@includegraphics\includegraphics
\AtBeginDocument{\let\includegraphics\saved@includegraphics}
\renewenvironment*{figure}{\@float{figure}}{\end@float}
\makeatother

\begin{document}

\title{Conformational Heterogeneity in Human Interphase Chromosome Organization Reconciles the FISH and Hi-C Paradox}

\author{Guang Shi}
\affiliation{Biophysics Program, Institute for Physical Science and Technology,\\
University of Maryland, College Park, MD, United States, 20742}
\author{D. Thirumalai}
\email{dave.thirumalai@gmail.com}
\affiliation{Department of Chemistry, University of Texas at Austin, Austin, TX, United States, 78712}

\begin{abstract}
Hi-C experiments are used to infer the contact probabilities between loci separated by varying genome lengths. Contact probability should decrease as the spatial distance between two loci increases. However, studies comparing Hi-C and FISH data show that in some cases the distance between one pair of loci, with larger Hi-C readout, is paradoxically larger compared to another pair with a smaller value of the contact probability. Here, we show that the FISH-Hi-C paradox can be resolved using a theory based on a Generalized Rouse Model for Chromosomes (GRMC). The FISH-Hi-C paradox arises because the cell population is highly heterogeneous, which means that a given contact is present in only a fraction of cells. Insights from the GRMC is used to construct a theory, without any adjustable parameters, to extract the distribution of subpopulations from the FISH data, which quantitatively reproduces the Hi-C data. Our results show that heterogeneity is pervasive in genome organization at all length scales, reflecting large cell-to-cell variations.
\end{abstract}

\maketitle

\section*{Introduction}
Through remarkable Hi-C experiments \cite{LiebermanAiden2009,Dixon2012,Sexton2012,Jin2013,Dekker2013,rao20143d}, based on the Chromosome Conformation Capture (3C) technique \cite{dekker2002capturing}, indirect glimpses of how the genome in a number of species is organized is starting to emerge. Because chromosome lengths are extremely large, ranging from tens of million base pairs in yeast to billion base pairs in human cells, they have to fold into highly compact structures in order to be accommodated in the cell nucleus. This requires that loci that are well separated along the one-dimensional genome sequence be close in three-dimensional (3D) space, which is made possible by forming a large number of loops. The high throughput Hi-C technique and its variants  are used to infer the probability of genome-wide contact formation between loci. In order to determine the contact probabilities between various loci in a genome, Hi-C experiments are performed in an ensemble of millions of cells. The readouts of the Hi-C experiment are contact frequencies  between a large number of loci from instantaneous snapshots of each cell, which are then used to construct the contact maps (Hi-C maps). The contact map is a matrix (2D representation) in which the elements represent the probability of contact between two loci that are separated by a specified genomic distance. A high contact count between two loci means that they interact with each other more frequently compared to ones with low contact count. 

A complementary and potentially a more direct way to determine genome organization is to measure spatial distances between loci using a low throughput Fluorescence \textit{In Situ} Hybridization (FISH) technique \cite{wang2016spatial,Bintu2018}. In addition to providing 3D distances in fixed cells, recently developed CRISPR–dCas9 FISH can be used to assay the dynamic behavior of loci in real-time \cite{Chen2013,Ma2015,Ma2016}. However, due to the current limitation of the number of distinct color probes, this method provides distance distribution information for only a small number of loci. 

FISH and Hi-C, which are entirely different experimental techniques, provide data on different aspects of genome organization. As noted in recent reviews \cite{giorgetti2016closing,Fraser15MicrbiolMolbiolReview}, there are problems associated with each method. It is difficult to reconcile Hi-C and FISH data for the following reasons. In interpreting the Hi-C contact map, one makes the intuitive assumption that loci with high probability contact must also be spatially close. However, it has been demonstrated using Hi-C and FISH data on the same chromosome that high contact frequency does not always imply proximity in space \cite{giorgetti2016closing,fudenberg2017fish,Bickmore2013,williamson2014spatial}. It should be noted that in most cases, the Hi-C and FISH measurements agree very well \cite{wang2016spatial, Bintu2018, szabo2018tads, finn2019extensive}. However, from a purely theoretical perspective, even a single contradiction is intriguing if the experimental errors can be ruled out. An outcome of our theory is that the discordance between FISH and Hi-C data arises because of extensive heterogeneity, which is embodied by the presence of a variety of conformations adopted by chromosomes in each cell. There are a variety of reasons, including differing fixation conditions and presence of two or more subpopulation of cells in which the chromosomes are present in distinct conformations, which could give rise to the discordance between FISH and Hi-C data, as lucidly described recently \cite{giorgetti2016closing,Fraser15MicrbiolMolbiolReview}.  Contact between two loci could be a rare event, not present in all cells, which is captured in Hi-C experiment by performing an ensemble average. We show using a precisely solvable model that due to the absence of a contact between two specific loci in a number of cells, those with higher contact frequency could be spatially farther on an average than two others with lower contact frequency.  In contrast, the probability of contact formation using the FISH method can only be obtained if the tail (small distance) of the distance distribution between locus $i$ and $j$ can be accurately measured. For a variety of reasons, including the size of the probe and the signal strength, this  not altogether straightforward using FISH technique. Thus, in order to combine the data from the two powerful techniques, it is crucial to establish a theoretical basis with potential practical link, between the contact probability and average spatial distance.

Setting aside the conditions under which FISH and Hi-C are performed (see recommendations for comparing the results from the two techniques with minimum bias which are described elsewhere \cite{giorgetti2016closing}) insights into the discordance between the two methods, when they occur, can be obtained using polymer physics concepts. Recently, Fudenberg and Imakaev \cite{fudenberg2017fish} performed polymer simulations using a strong attractive energy between two labelled loci and a ten fold weaker interaction between two other loci that are separated by a similar genomic distance. In addition, they also reported simulations based on the loop extrusion model. Both these types of simulations  showed there could be discordance between FISH and Hi-C, which we refer to as the FISH-Hi-C paradox. However, they did not provide any solution to the paradox, which is the principle goal of this work. 

In addition, recent single-cell Hi-C \cite{Stevens2017, Flyamer2017,Tan2018} and FISH experiments \cite{wang2016spatial,szabo2018tads, Bintu2018, finn2019extensive} have revealed that there are substantial cell-to-cell variations on genome organization. However, how to utilize the data reported in these experiments to enhance our understanding of 3D genome structural heterogeneity has not been unexplored. One approach is to create an appropriate polymer model based on Hi-C and imaging data, which would readily allow us to probe the structural variability using simulations \cite{shukron2017transient, shi2018interphase, liu2018chain, Buckle2018}. Indeed, it has been shown, using Hi-C and FISH data as well simulations \cite{Buckle2018}, that if the conformation of the chromatin fiber is taken to be homogeneous then trends observed in the FISH data could not be predicted. However, using simulations and including two levels of chromatin organization (open and compact) qualitative trends observed in the FISH data could be recovered \cite{Buckle2018}.

Here, we first establish a relationship between the contact probability and the mean spatial distance using an analytically solvable Generalized Rouse Chromosome Model (GRMC), which incorporates the presence of CTCF/cohein mediated loops. The GRMC may be thought of as an ideal chromosome model, very much in the spirit of the Rouse model for polymers, in which conceptual issues such as the origin of the FISH-Hi-C paradox can be rigorously established. We first consider the solvable homogeneous limit in which contacts are present in all the cells. In this case, precise numerical and analytical results show that there is a simple relation between the contact probability, $P$, and the ensemble mean 3D distance $\langle R\rangle$. However, the unavoidable heterogeneity in the cell populations in Hi-C experiments, results in contacts between loci only in a fraction of cells. We first show that a direct consequence of the heterogeneity in both GRMC and chromosomes is that two loci ($m$ and $n$) that have higher probability ($P_{mn}$) of being in contact relative to another two loci ($k$ and $l$) does not imply a direct spatial correlation, a finding that has already been qualitatively established in previous studies \cite{giorgetti2016closing,fudenberg2017fish}. In other words, the average spatial distance between $m$ and $n$ ($\langle R_{mn}\rangle$) could be larger than $\langle R_{kl}\rangle$, the distance between loci $k$ and $l$, even if $P_{mn} > P_{kl}$. These results provide a basis for understanding the origin of FISH-Hi-C paradox.

We develop a fully theoretical approach, which allows us to provide quantitative insights into the extent of heterogeneity in chromosome organization. From our theory, it follows that the resolution of the FISH-Hi-C paradox requires invoking the notion of heterogeneity, which implies multiple populations of chromosomes coexist. By using the concepts that emerge from the study of the GRMC, we demonstrate that the information of cell subpopulations can be extracted by fitting the experimental FISH data using our theory, thus allowing us to calculate the Hi-C contact probabilities from the theoretically calculated cumulative distribution function of spatial distance (CDF) - a quantity that can be measured using FISH and super-resolution imaging methods. Our approach provides a theoretically based method to combine the available FISH and Hi-C data to produce a more refined characterization of the heterogeneous chromosome organization than is possible by using data from just one of the techniques. In other words, sparse data from both the experimental methods can be simultaneously harnessed to predict the 3D organization of chromosomes.

\section*{Results}

\textbf{Relating contact probability to mean spatial distance}. The exact relationship between $P_{mn}$ (contact probability between $m^{th}$ and $n^{th}$ locus) and the corresponding mean spatial distance, $\langle R_{mn}\rangle$ for GRMC (see Methods section for details of the derivation) is,

\begin{equation}\label{eq:1}
P_{mn}=\mathrm{erf}\bigg(\frac{2r_{\mathrm{c}}}{\sqrt{\pi}\langle R_{mn}\rangle}\bigg) - \frac{4}{\pi}\frac{r_{\mathrm{c}}}{\langle R_{mn}\rangle}e^{-\frac{4 r_{\mathrm{c}}^{2}}{\pi\langle R_{mn}\rangle^{2}}}\equiv R_{0}(\langle R_{mn}\rangle).
\end{equation}

\noindent The inverse of $R_{0}(\langle R_{mn}\rangle)$, the solution to Equation (\ref{eq:1}), gives the mean spatial distance $\langle R_{mn}\rangle$ as a function of the contact probability $P_{mn}$. Note that $m$ and $n$ are arbitrary locations of any two loci, and thus Equation (\ref{eq:1}) is general for any pair of loci. 

A couple of conclusions, relevant to the application to the chromosomes, follow from Equation (\ref{eq:1}).  (i)  Note that Equation (\ref{eq:1}) is an exact one-to-one relation between the mean distance $\langle R_{mn}\rangle$ and the contact probability $P_{mn}$ provided $r_\mathrm{c}$ is known, and if the contacts are present in all the cells, which is not the case in experiments. For small $P_{mn}$, it is easy to show from Equation (\ref{eq:1}) that $\langle R_{mn} \rangle \approx r_{\mathrm{c}}P_{mn}^{-1/3}$. For the ideal GRMC, this implies that for any $m,n,k,l$, if $P_{mn} < P_{kl}$ then $\langle R_{mn}\rangle > \langle R_{kl}\rangle$, a consequence anticipated on intuitive grounds. (ii) If the value of the contact probability $P$ and the threshold distance $r_{\mathrm{c}}$ are known precisely, then the distribution of the spatial distance can be readily computed by solving Equation (\ref{eq:1}) numerically. In Fig.1(b), we show the comparison between theory (Equation (\ref{eq:1})) and simulations (see Methods section for details). The simulated curves are computed as follows: first collect $(P_{mn},\langle R_{mn}\rangle)$ for every pair labeled $(m,n)$ where $P_{mn}$ and $\langle R_{mn}\rangle$ are computed using Equations (\ref{eq:17}-\ref{eq:18}) in the Methods section. The total number of pairs is $N(N-1)/2$. We then bin the points over the values of $P_{mn}$. Finally, the mean value of $\langle R_{mn}\rangle$ for each bin, $\langle R\rangle = E[\langle R_{mn}\rangle]$, is computed where $\mathrm{E}[\cdots]$ is the binned average, which is computed using $(1/N_{i})\sum_{j=1}^{N_{i}}\langle R_{mn}\rangle^{j}$ where $N_{i}$ is the number of points in the $i^{th}$ bin. The bin size, $\Delta$, is centered at $P_{mn}$, spanning $P_{mn}-\Delta/2 \leq P_{mn}\leq P_{mn}+\Delta/2$. Using this procedure, we find (Fig.1) that the theory and simulations are in perfect agreement, which validates the theoretical result.

\textbf{Contact distance $r_{\mathrm{c}}$ affects the inferred value of the spatial distance}. However, in practice, the elements $P_{mn}$ are measured with (unknown) statistical errors, and the value of the contact threshold $r_{\mathrm{c}}$ is only estimated. In the Hi-C experiments, contact probabilities and $r_\mathrm{c}$ by implication, are determined by a series of steps that start with  cross-linking spatially adjacent loci using formaldehyde, chopping the chromatin into fragments using restriction enzymes, ligating the fragments with biotin, followed by sequence matching using deep sequencing methods \cite{Dekker2013}. Because of the inherent stochasticity associated with the overall Hi-C scheme, as well as the unavoidable heterogeneity (only a fraction of cells has a specific contact and the contact could be dynamic) in the cell population the relationship $P_{mn}$ and $\langle R_{mn}\rangle$ is not straightforward.

To illustrate how the uncertainty in $r_{\mathrm{c}}$ affects the determination of the spatial distance in GRMC even when the population is homogeneous (all cells have a specific contact), we plot the distributions of distance for $r_{\mathrm{c}}=0.02,0.03\mathrm{\ }\mu\mathrm{m}$ in Fig.1(c). A small change in $r_{\mathrm{c}}$ (from 0.02 $\mu\mathrm{m}$ to 0.03 $\mu\mathrm{m}$) completely alters the distance distribution $P(R)$, and hence the mean spatial distance (from $\approx 0.2\mu m $ to $\approx 0.3\mu m$). For the exactly solvable GRMC, this can be explained by noting that $\langle R_{mn}\rangle \approx r_{\mathrm{c}} P_{mn}^{-1/3}$ for small $P_{mn}$. Because $P_{mn}$ appears in the denominator, any uncertainty in $r_{\mathrm{c}}$ is amplified by $P_{mn}$, especially when $P_{mn}$ is small.\newline

\noindent\textbf{Heterogeneity causes paradox between FISH and Hi-C}. The expectation that the contact probability should decrease as the mean distance between the loci increases, which is the case in the exactly solvable ideal GRMC ($P_{mn}\approx r_{\mathrm{c}} \langle R_{mn}\rangle^{-3}$), is sometimes violated when the experimental data \cite{rao20143d} is analyzed \cite{giorgetti2016closing,fudenberg2017fish}. The paradox is a consequence of heterogeneity due to the existence of more than one population of cells, which implies that in some fraction of cells, contact between two loci exists while in others it is absent. Each distinct population has its own statistics. For instance, the probability distribution of the spatial distance between the $m^{th}$ and the $n^{th}$ loci, $P_{i,mn}(r)$, for one population of cells could be different from another population of cells $P_{j,mn}(r)$ where $i$ and $j$ are the indices for the two different populations (Fig.2(a)). The Hi-C experiments yield only an average value of the contact probability. Let us illustrate the consequence of the inevitable heterogeneous mixture of cell populations by considering the simplest case in which only two distinct populations, one with probability $\eta$ and the other $1-\eta$, are present (a generalization is presented below). For instance, in one population of cells, there is a CTCF loop between $m$ and $n$, and it is absent in the other population. The probability distribution of spatial distance between the $m^{th}$ and the $n^{th}$ loci is a superposition of distributions for each population. Using Equation \ref{eq:10}, the mixed distribution can be written as,

\begin{equation}\label{eq:2}
P(R_{mn}=r)=\sqrt{\frac{2}{\pi}}\bigg(\eta\frac{r^{2}}{\sigma^{3}_{1,mn}}e^{-\frac{r^{2}}{2\sigma^{2}_{1,mn}}}+(1-\eta)\frac{r^{2}}{\sigma^{3}_{2,mn}}e^{-\frac{r^{2}}{2\sigma^{2}_{2,mn}}}\bigg)
\end{equation}

\noindent where $\sigma_{1,mn}$ and $\sigma_{2,mn}$ are the parameters with different values characterizing the two populations. In the GRMC, $\sigma_{1,mn}$ and $\sigma_{2,mn}$ are related to the mean spatial distances in the two populations by $\langle R_{1,mn}\rangle=2\sqrt{2/\pi}\sigma_{1,mn}$ and $\langle R_{2,mn}\rangle=2\sqrt{2/\pi}\sigma_{2,mn}$. The mean spatial distance is, $\langle R_{mn}\rangle=\eta\langle R_{1,mn}\rangle+(1-\eta)\langle R_{2,mn}\rangle$, and the contact probability is $P_{mn}=\eta P_{1,mn} + (1-\eta)P_{2,mn}$ where $P_{1,mn}$ and $P_{2,mn}$ are the contact probabilities for each population, given by Equation (\ref{eq:1}), which depends on the values of $\langle R_{1,mn}\rangle$ and $\langle R_{2,mn}\rangle$ as well as $r_{\mathrm{c}}$.

If the values of $\langle R_{1,mn}\rangle$ and $\langle R_{2,mn}\rangle$ are unknown (as is the case in Hi-C experiments), and only the value of the contact probability between the two loci is provided, one can not uniquely determine the values of the mean spatial distances. This is the origin of the Hi-C and FISH data paradox. In Figs.2(b)-2(e) we show an example of the paradox for a particular set of parameters ($\eta,\sigma_{1,mn}, \sigma_{2,mn}$). Pair \#1 has a larger contact probability than pair \#2, while also exhibiting a larger mean spatial distance. The GRMC explains in simple terms the origin of the paradox.

To systematically explore the parameter space, we display $\langle R_{mn}\rangle$ and $P_{mn}$ as heat maps showing $\langle R\rangle_{1,mn}$ versus $\langle R\rangle_{2,mn}$ for different values of $\eta$ (Fig.3). When there is a single homogenous population ($\eta=0.0$), the mean spatial distance $\langle R_{mn}\rangle$ and contact probability $P_{mn}$ depend only on the value of $\langle R_{2,mn}\rangle$ (upper panel in Fig.3). In this case, there is a precise one-to-one mapping between $\langle R_{mn}\rangle$ and $P_{mn}$. However, if $\eta \ne 0$ ($\eta=0.3$, lower panel in Fig.3) then the relation between $P_{mn}$ and $\langle R_{mn}\rangle$ is complicated. The contour lines for $P_{mn}$ cross the contour lines of $\langle R_{mn}\rangle$, which implies that for a given value of $P_{mn}$, one cannot infer the value of $\langle R_{mn}\rangle$ without knowing the value of $\eta$, $\langle R_{1,mn}\rangle$ and $\langle R_{2,mn}\rangle$. For instance, the triangle and circle shown for $\eta=0.3$ in Fig.3 demonstrate an example of the paradox in which $\langle R(\blacktriangledown)\rangle(=57a) > \langle R(\bullet)\rangle (=40a)$ whereas $P(\blacktriangledown) (\approx 7.7\times 10^{-4}) > P(\bullet) (\approx 3.9\times 10^{-4})$.\newline

\noindent\textbf{Extracting cell subpopulation information from FISH data}. Can we extract the information about subpopulations from experimental data so that the result from two vastly different techniques can be reconciled? To answer this question, we first generalize our theory derived from GRMC to real chromatins. The generalization of Equation (\ref{eq:2}) is, 

\begin{equation}\label{eq:3}
P(R_{mn}=r)=\eta P(r|\langle R_{1,mn}\rangle)+(1-\eta)P(r|\langle R_{2,mn}\rangle)
\end{equation}

\noindent where $P(r|\langle R_{1,mn}\rangle)$ and $P(r|\langle R_{2,mn}\rangle)$ are the Redner-des Cloizeaux distribution of distances for polymers \cite{jannink1990polymers,redner1980distribution} (Supplementary Note 1 and Supplementary Figure 1). The distribution $P(r|\langle R_{mn}\rangle)$ is rigorously known for self-avoiding homopolymer in a good solvent, generalized Rouse model (Equation \ref{eq:10} in the Methods section), and a semi-flexible polymer \cite{Hyeon2006,Wilhelm1996}. However, a simple analytic expression for chromosomes is not known. By assuming that the Redner-des Cloizeaux form for $P(r|\langle R_{mn}\rangle)$ also holds for chromosomes (see Supplementary Equation 1 for details), we find that $g=1$ and $\delta=5/4$ in Supplementary Equation 1. These parameters were previously extracted using experimental data \cite{wang2016spatial}, and the Chromosome Copolymer Model (CCM) for chromosomes \cite{shi2018interphase}. The value of $g$ is inferred from the scaling relationship between mean spatial distance $\langle R\rangle$ and contact probability $P$,  $P\sim \langle R\rangle ^{3+g}$. The value of $\delta$ is computed as $\delta=1/(1-\nu)$. $\nu$ is inferred from scaling $\langle R(s)\rangle\sim s^{\nu}$ where $s$ is the genomic distance.

The integral of Equation (\ref{eq:3}) up to $R$, which is the cumulative distribution function $\mathrm{CDF}(R)$, can be used to fit the FISH data. Thus, the probability of contact formation can be computed as, $\int_{0}^{r_{\mathrm{c}}}P(r|\langle R\rangle)\mathrm{d}r$ where $r_\mathrm{c}$ is the contact threshold. Using the data in \cite{rao20143d}, the $\mathrm{CDF}(R)$ for two pairs of loci are shown in Fig.4(a). By fitting the two experimentally measured curves to the theoretical prediction (see Supplementary Note 2), we obtain $\eta\approx0.42$ for peak4-loop and $\eta\approx 0.97$ for peak3-control. The parameters obtained can then be used to compute the contact probability. Since the Hi-C experiments measure the number of contact events instead of contact probability and the value of $r_{\mathrm{c}}$ is unknown, we compare the relative contact frequency, which is computed as $P_{i}/\langle P \rangle$ where $P_{i}$ is the contact probability computed using the model or the contact number measured in Hi-C for the $i^{th}$ pair and $\langle P\rangle$ is the mean value for all the pairs considered. First, we fit all the eight $\mathrm{CDF}(R)$ curves in \cite{rao20143d}.  the excellent agreement between theory and experiments is vividly illustrated in Supplementary Figure 2 and also manifested by the Kolmogorov-Smirnov statistics (Supplementary Note 5 and Supplementary Table 1). Second, we calculate their corresponding relative contact frequency (Fig.4(b)). Comparison of the theoretical calculations with Hi-C measurements shows excellent agreement (Fig.4(b)) with the Pearson correlation coefficient being 0.87. The contact probability is computed using $r_{\mathrm{c}}=10\mathrm{\ nm}$. Note that any value of $r_{\mathrm{c}}\leq 10\mathrm{\ nm}$ gives similar results (Supplementary Figure 3). The goodness of fits using different sets of $g$ and $\delta$ is summarized in Supplementary Table 2. The set of $g=0$ and $\delta=2$ gives equivalent good fits as the set of $g=1$ and $\delta=5/4$. It is also important to note that fitting the FISH data with the assumption that cell population is homogeneous leads to unphysical values of $g$ and $\delta$ and the Kolmogorov-Smirnov statistics are inferior (see Supplementary Note 4, Supplementary Figure 5 and Supplementary Table 3). 

Interestingly, the values of $\langle R_{1}\rangle$ obtained from fitting the four CTCF/cohesin mediated loops (peak(1,2,3,4)-loop) are all about $0.25-0.35\mu\mathrm{m}$ ($R_{1,\mathrm{peak1-loop}}\approx 0.24\mu m$,  $R_{1,\mathrm{peak2-loop}}\approx 0.33\mu m$, $R_{1,\mathrm{peak3-loop}}\approx 0.35\mu m$,$R_{1,\mathrm{peak4-loop}}\approx 0.30\mu m$) regardless of their genomic separation (see Supplementary Table 1), suggesting that the mechanism of looping between CTCF motifs are similar with a mean spatial distance $\approx 0.3\mathrm{\ \mu m}$.  The physically reasonable value of $\langle R_{mn}\rangle \approx 0.3 \mathrm{\nu m}$ for all peak-loop pairs shows that these CTCF-mediated contacts describe molecular interactions between loci that are separated by a few hundred kilo base pairs. It has been shown that these contacts, referred to as peaks \cite{rao20143d} are significantly closer in space than others that are separated by similar genomic distance. The peak-loop contacts correspond to chromatin loops with the loci in the peaks being the anchor points between a specific loop. In sharp contrast, the distances between peak$i$-control ($i$ goes from 1 to 4), which are greater than the distances between peak loci,  vary ranging from $\approx 0.47 \mathrm{\ \mu m}$ to $\approx 0.67 \mathrm{\ \mu m}$ (see Supplementary Table 1). It is likely that these contacts are more dynamic because they are not be anchored by CTCF binding proteins.   \newline

\noindent\textbf{Massive heterogeneity in chromosome organization}. In a recent study \cite{finn2019extensive}, which combined Hi-C and high-throughput optical imaging to map contacts within single chromosomes in human fibroblasts, revealed massive heterogeneity. Such extensive existence of a large number of conformations, leading to multiple or nearly continuous distribution of subpopulations, was much greater than previously anticipated. Although, the results in Fig.4 quantitatively reveal heterogeneity associated with CTCF loops by considering only two dominant subpopulations, the most recent experiment requires a generalization of the theory. In principle, our theory also applies to interactions of any nature, not only the CTCF loops. In doing so, it may be more reasonable to assume a continuous distribution of subpopulations, $P(\langle R\rangle)$, (see Supplementary Note 6, Supplementary Note 7, and Supplementary Figure 6 for generalization) instead of two discrete subpopulations, $\langle R_1\rangle$ and $\langle R_2\rangle$, which of course is much simpler and may suffice in many cases as the results in Fig. 4 illustrate. As a proof of concept of our theory, we solve $P(\langle R\rangle)$ for the eight pairs of contacts analyzed in the previous section. The results are shown in Supplementary Figure 7. In all cases, $P(\langle R\rangle)$s are found to be multi-modal. For peak1/2/3/4-control and peak3-loop, $P(\langle R\rangle)$ yield peaks located at positions very close to $\langle R_1\rangle$ and $\langle R_2\rangle$ shown in Supplementary Table 1, justifying the effectiveness of the theory. To show that our theory has a broader range of applicability, we use the FISH data from the recent study \cite{finn2019extensive}, which reports spatial distance measurements for 212 pairs of loci.  $P(\langle R\rangle)$ is solved for each of a total of 212 pairs of loci. To illustrate our results, we compare in Fig.5 the predicted  $\mathrm{CDF}(r)$  and the experimentally measured $\mathrm{CDF}(r)$, as well as the $P(\langle R\rangle)$ obtained by fitting for six pairs of loci as examples in Fig.5. The results show substantial variations in $\langle R\rangle$, manifested by the multiple peaks and wide spreaded variations in $P(\langle R\rangle)$. Remarkably, the calculated $\mathrm{CDF}(r)$ (without any adjustable parameters) and the measured $\mathrm{CDF}(r)$ are in excellent agreement for the six loci pairs, which were arbitrarily chosen for illustration purposes. The residual errors between the two, shown as insets in Fig.5, are extremely small. 

In Fig. 6a we show the normalized distributions $P(\langle R\rangle/\mu(\langle R\rangle))$ for each of the 212 pairs of loci (see Supplementary Figure 8 for each pair as a separate figure). We expect that $P(\langle R\rangle/\mu(\langle R\rangle))$ should be narrowly distributed around value 1 if there is only one population. However, many $P(\langle R\rangle/\mu(\langle R\rangle))$ show multiple peaks with large variations. To further quantify the extent of heterogeneity, we calculate the coefficient of variation, $\mathrm{CV} = \sigma(\langle R\rangle)/\mu (\langle R\rangle)$ where $\sigma(\langle R\rangle)$ and $\mu(\langle R\rangle)$ are the standard deviation and the mean of $\langle R\rangle$, respectively. If there is only one population associated with $\langle R\rangle$, $\mathrm{CV}$ should have a value of around zero. Fig. 6b shows the histogram of $\mathrm{CV}$ for all 212 pairs of loci. The $\mathrm{CV}$ values are widely distributed, suggesting that 3D structural heterogeneity is common and is associated with many pairs of loci rather than a few. Thus, the analyses of experimental data are not possible without taking heterogeneity into account. The theory presented here is sufficiently general and simple that it can be used to calculate the measurable quantities readily.

\noindent\textbf{The role of loop extrusion in chromosome heterogeneity}. What is the origin of heterogeneity in the individual cell populations? There are two possibilities. The first one is static heterogeneity: each subpopulation explores a distinct region of the genomic folding landscape (GFL) (Fig.7a). The second is the dynamic heterogeneity. Each cell explores a local minimum of the GFL before transiting to another local minimum (Fig.7b). The only assumption in the application of our theory to genome organization is that there must be more than one population of cells, which does not violate the observation that the Hi-C experiment report only the average contact probability over millions of cells. Dynamic looping would be an example of the dynamic heterogeneity where the CTCF/cohein mediated loops are formed and broken dynamically on a fast time scale compared to the life time of a cell. Such a picture is supported by recent single-cell molecule experiments \cite{Hansen2017elife,Hansen2017nucleus}. The average residence time of CTCF/cohesin complex is shown to be in the range of a few to tens of minutes, which is much smaller compared to the time scale of the cell cycle (15-30 hours). Loop extrusion model \cite{alipour2012self,sanborn2015chromatin,fudenberg2016formation} is another possible origin of dynamic heterogeneity. In the loop extrusion model, it is thought that cohesins extrude loops along the chromsome fiber, which could detach stochastically. At any given time, there would be many subpopulations, each characterized by a distinct set of loops in the chromosome. Indeed, our analyses of the most recent high throughput optical imaging data lend credence to the notion that mutiple subpopulations in chromosomes arise because of massive dynamic heterogeneity. Our theory also gives an indirect theoretical justification for the work in \cite{fudenberg2017fish} in which the authors found the loop extrusion model could lead to the $[P_{mn},\langle R_{mn}\rangle]$ paradox.

Single-cell temporal information is necessary to determine whether the loops are static or dynamic or a combination of the two (Fig.7c). Hence, the combination of the dynamic FISH technique such as CRISPR-dCas9 FISH and single-cell Hi-C would be crucial for us to fully understand the organization of genomes. Our theory provides a theoretically rigorous method based on polymer physics to connect the results from measurements using the two vastly different techniques.

\section*{Discussion}
From polymer physics for single chains it follows that in a homogeneous system, the contact probability and mean 3D distances are linked, resulting in a power-law relation connecting the two quantities that can be measured using Hi-C and FISH techniques. However, the one-to-one mapping does not hold in Hi-C experiments because of the presence of a mixture of distinct cell subpopulations each characterized by its own statistics leads to heterogeneity, which in turn gives rise to the [$P_{mn}$,$\langle R_{mn}\rangle$] paradox. We show that the theory based on precisely solvable GRMC could be used to solve  the paradox in practice.  The theory can be readily used to analyze data from experiments, provided the FISH and Hi-C experiments are done under similar conditions \cite{rao20143d}. The central result of the theory in Equation (\ref{eq:3}) can be used to analyze the available sparse FISH data. We show that the fraction of cell subpopulations ($\eta$ in Equation (\ref{eq:3})) and the generalization derived in Supplementary Note 6 can be extracted by fitting the FISH data using our theory. From Equation (\ref{eq:3}) we calculate the Hi-C contact probabilities, thus establishing that the theory resolves the [$P_{mn}$,$\langle R_{mn}\rangle$] paradox. 

In this work, we confine ourselves to two-point interactions, which allows us to consider one pair of loci at a time. However, recent experiments probing multi-point interactions have suggested that formations of loops are likely to be cooperative  \cite{quinodoz2018higher, Bintu2018}, such that the formation of one loop could facilitate the formation of a nearby loop. Such cooperative loop formation was previously shown in an entirely different context involving the folding of proteins directed by disulfide bond formation \cite{Camacho95PNAS}. It can be shown within our framework that the formation of one loop can certainly increase the probability of formation of another loop. The theoretical basis for this finding is given in the Supplementary Note 8.

The reconciliation of the FISH and Hi-C data using polymer physics concepts is the first key step in integrating the data from these experimental techniques to construct the 3D structures of chromosomes. The work described here provides a theoretical basis for accomplishing this important task. Finally, our results suggest that heterogeneity in contact formation is an intrinsic property of genome organization, and hence the acquisition of single-cell experimental data is crucial to further our understanding of both the dynamics and the heterogeneous structural organization of chromosomes.\newline

\section*{Methods}
\textbf{Generalized Rouse Model for Chromosome}.
In order to derive an approximate relationship connecting contact probabilities between loci and the three dimensional distances, we use a variant of the random loop model \cite{bryngelson1996internal,shukron2017statistics}. We first consider a minimal cross-linked phantom chain model, which incorporates the presence of CTCF/cohein mediated loops \cite{rao20143d}. The model, originally introduced for describing physical gels \cite{bryngelson1996internal}, and more recently used for chromosome dynamics in a number of insightful studies \cite{shukron2017statistics,shukron2017transient}, could be viewed as a Generalized Rouse Chromosome Model (GRMC) \cite{hyeon2008force,o2009accurate}. The cross-links modeling the CTCF/cohein mediated loops here are not random. Their locations are predetermined by the Hi-C data \cite{rao20143d}. 

The equations of motion for the GRMC is \cite{doi1988theory},
\begin{equation}\label{eq:4}
\xi\frac{\mathrm{d}\boldsymbol{\mathrm{R}}}{\mathrm{d}t} = \boldsymbol{\mathrm{A}}\boldsymbol{\mathrm{R}}+\boldsymbol{\mathrm{F}}
\end{equation}
\noindent where $\xi$ is the friction coefficient, $\boldsymbol{\mathrm{R}}=[\boldsymbol{\mathrm{r}}_{1},\boldsymbol{\mathrm{r}}_{2},...,\boldsymbol{\mathrm{r}}_{N}]^{\mathrm{T}}$ with $\boldsymbol{\mathrm{r}}_{i}$ being the position of the $i^{th}$ locus. The vector $\boldsymbol{\mathrm{F}}=[\boldsymbol{\mathrm{f}}_{1},\boldsymbol{\mathrm{f}}_{2},...,\boldsymbol{\mathrm{f}}_{N}]^{\mathrm{T}}$ ($\mathrm{T}$ is the transpose), where $\boldsymbol{\mathrm{f}}_{i}$ is the Gaussian random force acting on the $i^{th}$ locus, characterized by $\langle f_{n}(t) \rangle$=0 and $\langle f_{n\alpha}(t)f_{m\beta}(t^{\prime}) \rangle=2\xi k_{\mathrm{B}}T\delta_{nm}\delta_{\alpha\beta}\delta(t-t^{\prime})$; $\boldsymbol{\mathrm{A}}$ is the $N\times N$ connectivity matrix, embedding the information of chain connectivity and the location of the loops connecting two loci (Fig. 1(a));

\begin{equation}\label{eq:5}
    A_{mn}= 
\begin{dcases}
    -2\kappa-|\Sigma_{m}|\omega,& \text{if } m = n\neq 1\text{\ or\ }N\\
    -\kappa - |\Sigma_{m}|\omega,& \text{if } m = n =1\text{ or }N\\
    \kappa,              & \text{if } |m-n|=1\\
    \omega,             & \text{if } |m-n|>1,\text{and connected in }\Sigma\\
    0,                       & \text{if otherwise} 
\end{dcases}
\end{equation}

\noindent where $\Sigma$ is the set of indices representing the loci pairs specifying the CTCF facilitated loop anchors, and $|\Sigma_{m}|$ is the number of loops connected to the $m^{th}$ locus. The spring constant $\kappa$ enforces chain connectivity, and $\omega$ is the associated spring constant for a CTCF pair. Note that the GRMC model does not account for excluded volume interactions, which in the modeling of chromatin is often justified by noting that topoisomerases enable chain crossing. Our purpose is to use GRMC to first illustrate concretely the challenges in going from the measured average contact map to spatial organization, precisely. More importantly, using the insights from the study of the GRMC, we provide a solution to  the FISH-Hi-C paradox. 

Since $\boldsymbol{\mathrm{A}}$ in Equation (\ref{eq:5}) is a real symmetric matrix, it can be diagonalized using the orthonormal matrix $\boldsymbol{\mathrm{V}}$, 
\begin{equation}\label{eq:6}
\boldsymbol{\mathrm{V}} \boldsymbol{\mathrm{A}} \boldsymbol{\mathrm{V}}^{T}=\boldsymbol{\mathrm{\Lambda}}=\mathrm{diag}(\lambda_{0},\lambda_{1},...,\lambda_{N-1})
\end{equation}
\noindent where $\lambda_{0},\lambda_{1},...,\lambda_{N-1}$ are the eigenvalues of $\boldsymbol{\mathrm{A}}$. By defining $\boldsymbol{\mathrm{X}}=\boldsymbol{\mathrm{V}}\boldsymbol{\mathrm{R}}$ and using $\boldsymbol{\mathrm{R}}=\boldsymbol{\mathrm{V}}^{T}\boldsymbol{\mathrm{X}}$ and $\boldsymbol{\mathrm{V}}\boldsymbol{\mathrm{V}}^{T}=\boldsymbol{\mathrm{I}}$, we obtain the equations of motion of the normal coordinates $\boldsymbol{\mathrm{X}}$,

\begin{equation} \label{eq:7}
\xi\frac{\mathrm{d}\boldsymbol{\mathrm{X}}}{\mathrm{d}t} = \boldsymbol{\mathrm{\Lambda}}\boldsymbol{\mathrm{X}}+\boldsymbol{\mathrm{f}}.
\end{equation}
\noindent Because $\boldsymbol{\mathrm{\Lambda}}$ is a diagonal matrix, the normal coordinates of the GRMC $\boldsymbol{\mathrm{X}}_{p}$ are decoupled. Using the normal modes, $\boldsymbol{\mathrm{X}}$, the physical quantities associated with the polymer can be readily calculated. Therefore, for GRMC with a predetermined set of CTCF/cohein mediated loops, we can solve for the eigenvalues of the connectivity matrix $\boldsymbol{\mathrm{A}}$, and the orthonormal matrix $\boldsymbol{\mathrm{V}}$ numerically, and thus calculate the contact probability and spatial distance precisely.

\textbf{Relation between contact probability and mean spatial distance}.
The vector between the positions of the $m^{th}$ and the $n^{th}$ loci may be written as,

\begin{equation}\label{eq:8}
\boldsymbol{\mathrm{R}}_{m}-\boldsymbol{\mathrm{R}}_{n}=\sum_{p=0}^{N-1}(V_{pm}-V_{pn})\boldsymbol{\mathrm{X}}_{p}
\end{equation}

\noindent where $V_{pm}$ and $V_{pn}$ are the elements of orthonormal matrix $\boldsymbol{\mathrm{V}}$. The equilibrium solution of Equation (\ref{eq:7}) yields, $\lim_{t\to\infty}X_{p,\alpha}(t)\sim\mathcal{N}(0,-\frac{k_{\mathrm{B}}T}{\lambda_{p}})$, where $\alpha=x,y,z$, $\mathcal{N}$ is Gaussian distribution. Therefore,

\begin{equation}\label{eq:9}
\lim_{t\to\infty}R_{mn,\alpha}(t)\sim\mathcal{N}(0,-\sum_{p=0}^{N-1}(V_{pm}-V_{pn})^{2}\frac{k_{\mathrm{B}}T}{\lambda_{p}})\equiv\mathcal{N}(0,\sigma_{mn,\alpha}^{2}).
\end{equation}

\noindent where $\sigma_{mn,\alpha}=-\sum_{p=0}^{N-1}(V_{pm}-V_{pn})^{2}(k_{\mathrm{B}}T/\lambda_{p})$. Since the model is isotropic, it follows that $\sigma^{2}_{mn,x}=\sigma^{2}_{mn,y}=\sigma^{2}_{mn,z}\equiv\sigma^{2}_{mn}$. The mean distance $\langle R_{mn}\rangle$ is related to $\sigma_{mn}$ through $\langle R_{mn}\rangle=2\sqrt{2/\pi}\sigma_{mn}$. The distribution of distance between the $m^{th}$ and the $n^{th}$ loci, $\lim_{t\to\infty}|\boldsymbol{\mathrm{R}}_{mn}(t)|=\lim_{t\to\infty}\sqrt{\sum_{\alpha}R^{2}_{mn,\alpha}(t)}$ is a non-central chi distribution (we will neglect the notation $\lim_{t\to\infty}$ from now on),

\begin{equation}\label{eq:10}
P(R_{mn}=r)=\sqrt{\frac{2}{\pi}}\frac{1}{\sigma_{mn}}e^{-r^{2}/(2\sigma^{2}_{mn})}\frac{r^{2}}{\sigma^{2}_{mn}}.
\end{equation}

\noindent The contact probability $P_{mn}$, for a given threshold $r_{\mathrm{c}}$ (contact exists if $r\leq r_{\mathrm{c}}$), computed using Equation \ref{eq:10} yields, 

\begin{equation}\label{eq:11}
\begin{split}
P_{mn}&=\int_{0}^{r_{\mathrm{c}}}\mathrm{d}r\sqrt{\frac{2}{\pi}}\frac{1}{\sigma_{mn}}e^{-r^{2}/(2\sigma^{2}_{mn})}\frac{r^{2}}{\sigma^{2}_{mn}}\\
&=\mathrm{Erf}\bigg(\frac{r_{\mathrm{c}}}{\sqrt{2}\sigma_{mn}}\bigg) - \sqrt{\frac{2}{\pi}}e^{-\frac{r_{\mathrm{c}}^{2}}{2\sigma_{mn}^{2}}}\frac{r_{\mathrm{c}}}{\sigma_{mn}}.
\end{split}
\end{equation}  

\noindent The mean spatial distance $\langle R_{mn}\rangle$ is given by,

\begin{equation}\label{eq:12}
\langle R_{mn} \rangle = \int_{0}^{\infty}\mathrm{d}r r \sqrt{\frac{2}{\pi}}\frac{1}{\sigma_{mn}}e^{-r^{2}/(2\sigma^{2}_{mn})}\frac{r^{2}}{\sigma^{2}_{mn}}=2\sqrt{\frac{2}{\pi}}\sigma_{mn}.
\end{equation}

\noindent Using Equations (\ref{eq:11}) and (\ref{eq:12}), the desired relation between $P_{mn}$ and $\langle R_{mn}\rangle$ becomes,

\begin{equation}\label{eq:13}
P_{mn}=\mathrm{erf}\bigg(\frac{2r_{\mathrm{c}}}{\sqrt{\pi}\langle R_{mn}\rangle}\bigg) - \frac{4}{\pi}\frac{r_{\mathrm{c}}}{\langle R_{mn}\rangle}e^{-\frac{4 r_{\mathrm{c}}^{2}}{\pi\langle R_{mn}\rangle^{2}}}\equiv R_{0}(\langle R_{mn}\rangle).
\end{equation}

Equation \ref{eq:13} is identical with Equation (\ref{eq:1}) in the main text.

\textbf{Simulations}.
The energy function for the GRMC is,
\begin{equation} \label{eq:14}
U(\boldsymbol{\mathrm{r}}_{1},...,\boldsymbol{\mathrm{r}}_{N})=\sum_{i=1}^{N-1}U_{i}^{\mathrm{S}}+\sum_{\{p,q\}}U_{\{p,q\}}^{\mathrm{L}}.
\end{equation}
\noindent For the bonded stretch potential, $U_{i}^{\mathrm{S}}$, we use,
\begin{equation} \label{eq:15}
U_{i}^{\mathrm{S}}=\frac{\kappa}{2}(|\boldsymbol{\mathrm{r}}_{i+1} - \boldsymbol{\mathrm{r}}_{i}|-a)^{2},
\end{equation}
\noindent where $a$ is the equilibrium bond length. The interaction between the loop anchors is also modeled using a harmonic potential,
\begin{equation} \label{eq:16}
U_{\{p,q\}}^{\mathrm{L}}=\frac{\omega}{2}(|\boldsymbol{\mathrm{r}}_{p} - \boldsymbol{\mathrm{r}}_{q}| - a)^{2}
\end{equation}
\noindent where the spring constant is associated with the CTCF facilitated loops, and $\{p,q\}$ represent the indices of the loop anchors, which are taken from the Hi-C data \cite{rao20143d} (Supplementary Note 3).  We simulate the chromosome segment from 146 Mbps to 158 Mbps of Chromosome 5. Each monomer represents 1200 bps, resulting total number of coarse-grained loci $N=10,000$. 

In order to accelerate conformational sampling, we perform Langevin Dynamics simulations at low friction \cite{honeycutt1992nature}. We simulate each trajectory for $10^{8}$ time steps, and save the snapshots every $10,000$ time steps. We generate ten independent trajectories, which are sufficient to obtain reliable statistics (Supplementary Figure 4).

\textbf{Data analyses}.
The contact probability between the $m^{th}$ and $n^{th}$ loci in the simulation is calculated using,

\begin{equation}\label{eq:17}
P_{mn} = \frac{1}{TM}\sum_{a=1}^{M}\sum_{t=1}^{T}\Theta(r_{\mathrm{c}}-|\boldsymbol{\mathrm{r}}^{(a)}_{m}(t)-\boldsymbol{\mathrm{r}}^{(a)}_{n}(t)|),
\end{equation}

\noindent where $\Theta(\cdot)$ is the Heaviside step function, $r_{\mathrm{c}}$ is the threshold distance for determining the contacts, the summation is over the snapshots along the trajectory, and the total $M$ number of independent trajectories, and $T$ is the number of snapshots for a single trajectory. The mean spatial distance between the $i^{th}$ and the $j^{th}$ loci in the simulation is calculated using,

\begin{equation}\label{eq:18}
\langle R_{mn}\rangle = \frac{1}{TM} \sum_{a=1}^{M}\sum_{t=1}^{T}|\boldsymbol{\mathrm{r}}^{(a)}_{m}(t) - \boldsymbol{\mathrm{r}}^{(a)}_{n}(t)|.
\end{equation}

\noindent The objective is to go from $P_{mn}$ to $\langle R_{mn}\rangle$, and to determine, if in doing so, we get reasonable results. Because these quantities can be computed precisely in the GRMC, the $[P_{mn}, \langle R_{mn}\rangle]$ relationship can be tested, which allows us to obtain the needed cues to solve the FISH-Hi-C paradox.

\textbf{Data availability}. All relevant data supporting the findings of this study are available within the article and its Supplementary Information files or upon requests from the corresponding author. The Hi-C and FISH experimental data used in this study are publicly available from GEO database under accession number GSE63525 [\url{https://www.ncbi.nlm.nih.gov/geo/query/acc.cgi?acc=GSE63525}] and from 4DN portal at \url{https://data.4dnucleome.org/publications/80007b23-7748-4492-9e49-c38400acbe60/}. The processed data is available upon request from the authors.

\textbf{Code availability}. The polymer simulations are performed using LAMMPS Molecular Dynamics Simulation software \cite{Plimpton1995}, which is an open source code available at \url{http://lammps.sandia.gov}. The codes used to analyze data in the present study are deposited to Github repository \url{https://github.com/anyuzx/chromosome-heterogeneity-analysis}.


\begin{thebibliography}{}
\expandafter\ifx\csname url\endcsname\relax
  \def\url#1{\texttt{#1}}\fi
\expandafter\ifx\csname urlprefix\endcsname\relax\def\urlprefix{URL }\fi
\providecommand{\bibinfo}[2]{#2}
\providecommand{\eprint}[2][]{\url{#2}}

\end{thebibliography}


\begin{thebibliography}{10}
\section*{References}
\expandafter\ifx\csname url\endcsname\relax
  \def\url#1{\texttt{#1}}\fi
\expandafter\ifx\csname urlprefix\endcsname\relax\def\urlprefix{URL }\fi
\providecommand{\bibinfo}[2]{#2}
\providecommand{\eprint}[2][]{\url{#2}}

\bibitem{LiebermanAiden2009}
\bibinfo{author}{Lieberman-Aiden, E.} \emph{et~al.}
\newblock \bibinfo{title}{Comprehensive mapping of long-range interactions
  reveals folding principles of the human genome}.
\newblock \emph{\bibinfo{journal}{Science}} \textbf{\bibinfo{volume}{326}},
  \bibinfo{pages}{289--293} (\bibinfo{year}{2009}).

\bibitem{Dixon2012}
\bibinfo{author}{Dixon, J.~R.} \emph{et~al.}
\newblock \bibinfo{title}{Topological domains in mammalian genomes identified
  by analysis of chromatin interactions}.
\newblock \emph{\bibinfo{journal}{Nature}} \textbf{\bibinfo{volume}{485}},
  \bibinfo{pages}{376--380} (\bibinfo{year}{2012}).

\bibitem{Sexton2012}
\bibinfo{author}{Sexton, T.} \emph{et~al.}
\newblock \bibinfo{title}{Three-dimensional folding and functional organization
  principles of the drosophila genome}.
\newblock \emph{\bibinfo{journal}{Cell}} \textbf{\bibinfo{volume}{148}},
  \bibinfo{pages}{458--472} (\bibinfo{year}{2012}).

\bibitem{Jin2013}
\bibinfo{author}{Jin, F.} \emph{et~al.}
\newblock \bibinfo{title}{A high-resolution map of the three-dimensional
  chromatin interactome in human cells}.
\newblock \emph{\bibinfo{journal}{Nature}} \textbf{\bibinfo{volume}{503}},
  \bibinfo{pages}{290--294} (\bibinfo{year}{2013}).

\bibitem{Dekker2013}
\bibinfo{author}{Dekker, J.}, \bibinfo{author}{Marti-Renom, M.~A.} \&
  \bibinfo{author}{Mirny, L.~A.}
\newblock \bibinfo{title}{Exploring the three-dimensional organization of
  genomes: interpreting chromatin interaction data}.
\newblock \emph{\bibinfo{journal}{Nat. Rev. Genet.}}
  \textbf{\bibinfo{volume}{14}}, \bibinfo{pages}{390--403}
  (\bibinfo{year}{2013}).

\bibitem{rao20143d}
\bibinfo{author}{Rao, S.~S.} \emph{et~al.}
\newblock \bibinfo{title}{{A 3D map of the human genome at kilobase resolution
  reveals principles of chromatin looping}}.
\newblock \emph{\bibinfo{journal}{Cell}} \textbf{\bibinfo{volume}{159}},
  \bibinfo{pages}{1665--1680} (\bibinfo{year}{2014}).

\bibitem{dekker2002capturing}
\bibinfo{author}{Dekker, J.}, \bibinfo{author}{Rippe, K.},
  \bibinfo{author}{Dekker, M.} \& \bibinfo{author}{Kleckner, N.}
\newblock \bibinfo{title}{{Capturing chromosome conformation}}.
\newblock \emph{\bibinfo{journal}{Science}} \textbf{\bibinfo{volume}{295}},
  \bibinfo{pages}{1306--1311} (\bibinfo{year}{2002}).

\bibitem{wang2016spatial}
\bibinfo{author}{Wang, S.} \emph{et~al.}
\newblock \bibinfo{title}{{Spatial organization of chromatin domains and
  compartments in single chromosomes}}.
\newblock \emph{\bibinfo{journal}{Science}} \textbf{\bibinfo{volume}{353}},
  \bibinfo{pages}{598--602} (\bibinfo{year}{2016}).

\bibitem{Bintu2018}
\bibinfo{author}{Bintu, B.} \emph{et~al.}
\newblock \bibinfo{title}{Super-resolution chromatin tracing reveals domains
  and cooperative interactions in single cells}.
\newblock \emph{\bibinfo{journal}{Science}} \textbf{\bibinfo{volume}{362}}
  (\bibinfo{year}{2018}).

\bibitem{Chen2013}
\bibinfo{author}{Chen, B.} \emph{et~al.}
\newblock \bibinfo{title}{Dynamic imaging of genomic loci in living human cells
  by an optimized crispr/cas system}.
\newblock \emph{\bibinfo{journal}{Cell}} \textbf{\bibinfo{volume}{155}},
  \bibinfo{pages}{1479--1491} (\bibinfo{year}{2013}).

\bibitem{Ma2015}
\bibinfo{author}{Ma, H.} \emph{et~al.}
\newblock \bibinfo{title}{Multicolor {CRISPR} labeling of chromosomal loci in
  human cells}.
\newblock \emph{\bibinfo{journal}{Proc. Natl. Acad. Sci. U.S.A}}
  \textbf{\bibinfo{volume}{112}}, \bibinfo{pages}{3002--3007}
  (\bibinfo{year}{2015}).

\bibitem{Ma2016}
\bibinfo{author}{Ma, H.} \emph{et~al.}
\newblock \bibinfo{title}{Multiplexed labeling of genomic loci with {dCas}9 and
  engineered {sgRNAs} using {CRISPRainbow}}.
\newblock \emph{\bibinfo{journal}{Nat. Biotechnol.}}
  \textbf{\bibinfo{volume}{34}}, \bibinfo{pages}{528--530}
  (\bibinfo{year}{2016}).

\bibitem{giorgetti2016closing}
\bibinfo{author}{Giorgetti, L.} \& \bibinfo{author}{Heard, E.}
\newblock \bibinfo{title}{{Closing the loop: 3C versus} {DNA} {FISH}}.
\newblock \emph{\bibinfo{journal}{Genome Biol.}} \textbf{\bibinfo{volume}{17}}
  (\bibinfo{year}{2016}).

\bibitem{Fraser15MicrbiolMolbiolReview}
\bibinfo{author}{Fraser, J.}, \bibinfo{author}{Williamson, I.},
  \bibinfo{author}{Bickmore, W.~A.} \& \bibinfo{author}{Dostie, J.}
\newblock \bibinfo{title}{An overview of genome organization and how we got
  there: from {FISH} to {Hi-C}}.
\newblock \emph{\bibinfo{journal}{Microbiol. Mol. Biol. R.}}
  \textbf{\bibinfo{volume}{79}}, \bibinfo{pages}{347--372}
  (\bibinfo{year}{2015}).

\bibitem{fudenberg2017fish}
\bibinfo{author}{Fudenberg, G.} \& \bibinfo{author}{Imakaev, M.}
\newblock \bibinfo{title}{{FISH}-ing for captured contacts: towards reconciling
  {FISH} and {3C}}.
\newblock \emph{\bibinfo{journal}{Nat. Methods}} \textbf{\bibinfo{volume}{14}},
  \bibinfo{pages}{673--678} (\bibinfo{year}{2017}).

\bibitem{Bickmore2013}
\bibinfo{author}{Bickmore, W.~A.} \& \bibinfo{author}{van Steensel, B.}
\newblock \bibinfo{title}{Genome architecture: domain organization of
  interphase chromosomes}.
\newblock \emph{\bibinfo{journal}{Cell}} \textbf{\bibinfo{volume}{152}},
  \bibinfo{pages}{1270--1284} (\bibinfo{year}{2013}).

\bibitem{williamson2014spatial}
\bibinfo{author}{Williamson, I.} \emph{et~al.}
\newblock \bibinfo{title}{{Spatial genome organization: contrasting views from
  chromosome conformation capture and fluorescence in situ hybridization}}.
\newblock \emph{\bibinfo{journal}{Gene. Dev.}} \textbf{\bibinfo{volume}{28}},
  \bibinfo{pages}{2778--2791} (\bibinfo{year}{2014}).

\bibitem{szabo2018tads}
\bibinfo{author}{Szabo, Q.} \emph{et~al.}
\newblock \bibinfo{title}{{TADs} are {3D} structural units of higher-order
  chromosome organization in {Drosophila}}.
\newblock \emph{\bibinfo{journal}{Sci. Adv.}} \textbf{\bibinfo{volume}{4}},
  \bibinfo{pages}{eaar8082} (\bibinfo{year}{2018}).

\bibitem{finn2019extensive}
\bibinfo{author}{Finn, E.~H.} \emph{et~al.}
\newblock \bibinfo{title}{Extensive heterogeneity and intrinsic variation in
  spatial genome organization}.
\newblock \emph{\bibinfo{journal}{Cell}} \textbf{\bibinfo{volume}{176}},
  \bibinfo{pages}{1502--1515} (\bibinfo{year}{2019}).

\bibitem{Stevens2017}
\bibinfo{author}{Stevens, T.~J.} \emph{et~al.}
\newblock \bibinfo{title}{{3D structures of individual mammalian genomes
  studied by single-cell Hi-C}}.
\newblock \emph{\bibinfo{journal}{Nature}} \textbf{\bibinfo{volume}{544}},
  \bibinfo{pages}{59--64} (\bibinfo{year}{2017}).

\bibitem{Flyamer2017}
\bibinfo{author}{Flyamer, I.~M.} \emph{et~al.}
\newblock \bibinfo{title}{Single-nucleus hi-c reveals unique chromatin
  reorganization at oocyte-to-zygote transition}.
\newblock \emph{\bibinfo{journal}{Nature}} \textbf{\bibinfo{volume}{544}},
  \bibinfo{pages}{110--114} (\bibinfo{year}{2017}).

\bibitem{Tan2018}
\bibinfo{author}{Tan, L.}, \bibinfo{author}{Xing, D.}, \bibinfo{author}{Chang,
  C.-H.}, \bibinfo{author}{Li, H.} \& \bibinfo{author}{Xie, X.~S.}
\newblock \bibinfo{title}{Three-dimensional genome structures of single diploid
  human cells}.
\newblock \emph{\bibinfo{journal}{Science}} \textbf{\bibinfo{volume}{361}},
  \bibinfo{pages}{924--928} (\bibinfo{year}{2018}).

\bibitem{shukron2017transient}
\bibinfo{author}{Shukron, O.} \& \bibinfo{author}{Holcman, D.}
\newblock \bibinfo{title}{{Transient chromatin properties revealed by polymer
  models and stochastic simulations constructed from Chromosomal Capture
  data}}.
\newblock \emph{\bibinfo{journal}{PLoS Comput. Biol.}}
  \textbf{\bibinfo{volume}{13}}, \bibinfo{pages}{e1005469}
  (\bibinfo{year}{2017}).

\bibitem{shi2018interphase}
\bibinfo{author}{Shi, G.}, \bibinfo{author}{Liu, L.}, \bibinfo{author}{Hyeon,
  C.} \& \bibinfo{author}{Thirumalai, D.}
\newblock \bibinfo{title}{Interphase human chromosome exhibits out of
  equilibrium glassy dynamics}.
\newblock \emph{\bibinfo{journal}{Nat. Commun.}} \textbf{\bibinfo{volume}{9}},
  \bibinfo{pages}{3161} (\bibinfo{year}{2018}).

\bibitem{liu2018chain}
\bibinfo{author}{Liu, L.}, \bibinfo{author}{Shi, G.},
  \bibinfo{author}{Thirumalai, D.} \& \bibinfo{author}{Hyeon, C.}
\newblock \bibinfo{title}{Chain organization of human interphase chromosome
  determines the spatiotemporal dynamics of chromatin loci}.
\newblock \emph{\bibinfo{journal}{PLoS Comput. Biol.}}
  \textbf{\bibinfo{volume}{14}}, \bibinfo{pages}{e1006617}
  (\bibinfo{year}{2018}).

\bibitem{Buckle2018}
\bibinfo{author}{Buckle, A.}, \bibinfo{author}{Brackley, C.~A.},
  \bibinfo{author}{Boyle, S.}, \bibinfo{author}{Marenduzzo, D.} \&
  \bibinfo{author}{Gilbert, N.}
\newblock \bibinfo{title}{Polymer simulations of heteromorphic chromatin
  predict the 3d folding of complex genomic loci}.
\newblock \emph{\bibinfo{journal}{Mol. Cell}} \textbf{\bibinfo{volume}{72}},
  \bibinfo{pages}{786--797.e11} (\bibinfo{year}{2018}).

\bibitem{jannink1990polymers}
\bibinfo{author}{Jannink, G.} \& \bibinfo{author}{Des~Cloizeaux, J.}
\newblock \bibinfo{title}{Polymers in solution}.
\newblock \emph{\bibinfo{journal}{J. Phys-Condens. Mat.}}
  \textbf{\bibinfo{volume}{2}}, \bibinfo{pages}{1--24} (\bibinfo{year}{1990}).

\bibitem{redner1980distribution}
\bibinfo{author}{Redner, S.}
\newblock \bibinfo{title}{Distribution functions in the interior of polymer
  chains}.
\newblock \emph{\bibinfo{journal}{J. Phys. A-Math. Gen.}}
  \textbf{\bibinfo{volume}{13}}, \bibinfo{pages}{3525--3541}
  (\bibinfo{year}{1980}).

\bibitem{Hyeon2006}
\bibinfo{author}{Hyeon, C.} \& \bibinfo{author}{Thirumalai, D.}
\newblock \bibinfo{title}{Kinetics of interior loop formation in semiflexible
  chains}.
\newblock \emph{\bibinfo{journal}{J. Chem. Phys.}}
  \textbf{\bibinfo{volume}{124}}, \bibinfo{pages}{104905}
  (\bibinfo{year}{2006}).

\bibitem{Wilhelm1996}
\bibinfo{author}{Wilhelm, J.} \& \bibinfo{author}{Frey, E.}
\newblock \bibinfo{title}{Radial distribution function of semiflexible
  polymers}.
\newblock \emph{\bibinfo{journal}{Phys. Rev. Lett.}}
  \textbf{\bibinfo{volume}{77}}, \bibinfo{pages}{2581--2584}
  (\bibinfo{year}{1996}).

\bibitem{Hansen2017elife}
\bibinfo{author}{Hansen, A.~S.}, \bibinfo{author}{Pustova, I.},
  \bibinfo{author}{Cattoglio, C.}, \bibinfo{author}{Tjian, R.} \&
  \bibinfo{author}{Darzacq, X.}
\newblock \bibinfo{title}{{CTCF} and cohesin regulate chromatin loop stability
  with distinct dynamics}.
\newblock \emph{\bibinfo{journal}{{eLife}}} \textbf{\bibinfo{volume}{6}}
  (\bibinfo{year}{2017}).

\bibitem{Hansen2017nucleus}
\bibinfo{author}{Hansen, A.~S.}, \bibinfo{author}{Cattoglio, C.},
  \bibinfo{author}{Darzacq, X.} \& \bibinfo{author}{Tjian, R.}
\newblock \bibinfo{title}{Recent evidence that {TADs} and chromatin loops are
  dynamic structures}.
\newblock \emph{\bibinfo{journal}{Nucleus}} \textbf{\bibinfo{volume}{9}},
  \bibinfo{pages}{20--32} (\bibinfo{year}{2017}).

\bibitem{alipour2012self}
\bibinfo{author}{Alipour, E.} \& \bibinfo{author}{Marko, J.~F.}
\newblock \bibinfo{title}{{Self-organization of domain structures by
  DNA-loop-extruding enzymes}}.
\newblock \emph{\bibinfo{journal}{Nucleic Acids Res.}}
  \textbf{\bibinfo{volume}{40}}, \bibinfo{pages}{11202--11212}
  (\bibinfo{year}{2012}).

\bibitem{sanborn2015chromatin}
\bibinfo{author}{Sanborn, A.~L.} \emph{et~al.}
\newblock \bibinfo{title}{{Chromatin extrusion explains key features of loop
  and domain formation in wild-type and engineered genomes}}.
\newblock \emph{\bibinfo{journal}{Proc. Natl. Acad. Sci. U.S.A}}
  \textbf{\bibinfo{volume}{112}}, \bibinfo{pages}{E6456--E6465}
  (\bibinfo{year}{2015}).

\bibitem{fudenberg2016formation}
\bibinfo{author}{Fudenberg, G.} \emph{et~al.}
\newblock \bibinfo{title}{{Formation of chromosomal domains by loop
  extrusion}}.
\newblock \emph{\bibinfo{journal}{Cell Rep.}} \textbf{\bibinfo{volume}{15}},
  \bibinfo{pages}{2038--2049} (\bibinfo{year}{2016}).

\bibitem{quinodoz2018higher}
\bibinfo{author}{Quinodoz, S.~A.} \emph{et~al.}
\newblock \bibinfo{title}{{Higher-order inter-chromosomal hubs shape 3D genome
  organization in the nucleus}}.
\newblock \emph{\bibinfo{journal}{Cell}} \textbf{\bibinfo{volume}{174}},
  \bibinfo{pages}{744--757} (\bibinfo{year}{2018}).

\bibitem{Camacho95PNAS}
\bibinfo{author}{Camacho, C.~J.} \& \bibinfo{author}{Thirumalai, D.}
\newblock \bibinfo{title}{{Theoretical Predictions of Folding Pathways Using
  the Proximity Rule with Applications to BPTI}}.
\newblock \emph{\bibinfo{journal}{Proc. Natl. Acad. Sci. USA}}
  \textbf{\bibinfo{volume}{92}}, \bibinfo{pages}{1277--1281}
  (\bibinfo{year}{1995}).

\bibitem{bryngelson1996internal}
\bibinfo{author}{Bryngelson, J.} \& \bibinfo{author}{Thirumalai, D.}
\newblock \bibinfo{title}{{Internal constraints induce localization in an
  isolated polymer molecule}}.
\newblock \emph{\bibinfo{journal}{Phys. Rev. Lett.}}
  \textbf{\bibinfo{volume}{76}}, \bibinfo{pages}{542--545}
  (\bibinfo{year}{1996}).

\bibitem{shukron2017statistics}
\bibinfo{author}{Shukron, O.} \& \bibinfo{author}{Holcman, D.}
\newblock \bibinfo{title}{{Statistics of randomly cross-linked polymer models
  to interpret chromatin conformation capture data}}.
\newblock \emph{\bibinfo{journal}{Phys. Rev. E}} \textbf{\bibinfo{volume}{96}},
  \bibinfo{pages}{012503} (\bibinfo{year}{2017}).

\bibitem{hyeon2008force}
\bibinfo{author}{Hyeon, C.}, \bibinfo{author}{Morrison, G.} \&
  \bibinfo{author}{Thirumalai, D.}
\newblock \bibinfo{title}{{Force-dependent hopping rates of RNA hairpins can be
  estimated from accurate measurement of the folding landscapes}}.
\newblock \emph{\bibinfo{journal}{Proc. Natl. Acad. Sci. U.S.A}}
  \textbf{\bibinfo{volume}{105}}, \bibinfo{pages}{9604--9609}
  (\bibinfo{year}{2008}).

\bibitem{o2009accurate}
\bibinfo{author}{O’Brien, E.~P.}, \bibinfo{author}{Morrison, G.},
  \bibinfo{author}{Brooks, B.~R.} \& \bibinfo{author}{Thirumalai, D.}
\newblock \bibinfo{title}{{How accurate are polymer models in the analysis of
  F{\"o}rster resonance energy transfer experiments on proteins?}}
\newblock \emph{\bibinfo{journal}{J. Chem. Phys.}}
  \textbf{\bibinfo{volume}{130}}, \bibinfo{pages}{124903}
  (\bibinfo{year}{2009}).

\bibitem{doi1988theory}
\bibinfo{author}{Doi, M.} \& \bibinfo{author}{Edwards, S.~F.}
\newblock \emph{\bibinfo{title}{{The theory of polymer dynamics}}},
  vol.~\bibinfo{volume}{73} (\bibinfo{publisher}{oxford university press},
  \bibinfo{year}{1988}).

\bibitem{honeycutt1992nature}
\bibinfo{author}{Honeycutt, J.} \& \bibinfo{author}{Thirumalai, D.}
\newblock \bibinfo{title}{{The nature of folded states of globular proteins}}.
\newblock \emph{\bibinfo{journal}{Biopolymers}} \textbf{\bibinfo{volume}{32}},
  \bibinfo{pages}{695--709} (\bibinfo{year}{1992}).

\bibitem{Plimpton1995}
\bibinfo{author}{Plimpton, S.}
\newblock \bibinfo{title}{Fast parallel algorithms for short-range molecular
  dynamics}.
\newblock \emph{\bibinfo{journal}{J. Comput. Phys.}}
  \textbf{\bibinfo{volume}{117}}, \bibinfo{pages}{1--19}
  (\bibinfo{year}{1995}).

\end{thebibliography}

\clearpage

\section*{Acknowledgements:} We are grateful to the National Science Foundation (CHE 19-00093) and the Collie-Welch Regents Chair (F-0019) for supporting this work.

\section*{Conflict of interests:} The authors declare no competing interests.

\section*{Author contributions:} G. S. and D. T. designed and performed research, G.S. and D.T. analyzed data, G.S. and D.T. wrote the paper.
\clearpage

\begin{figure}[htb]
\centering
\includegraphics[width=\textwidth]{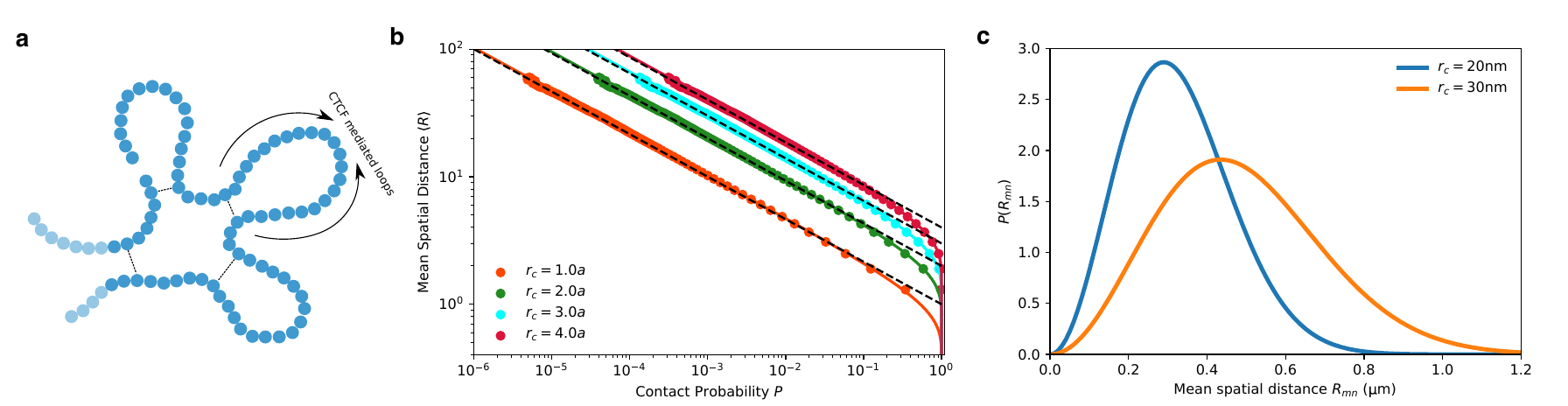}
\end{figure}
\captionof{figure}{Simulations demonstrate the power-law relation between contact probability and mean spatial distance and the effect of $r_{\mathrm{c}}$ on the inferred spatial distances. (\textbf{a}) A sketch of the Generalized Rouse Model for Chromosome (GRMC). Each bead represents a locus with a given resolution. Dashed lines represent harmonic bond between loop anchors. (\textbf{b}) Mean spatial distance $\langle R\rangle$ as a function of the contact probability $P$. The solid lines are obtained using Equation (\ref{eq:1}) for different values of $r_{\mathrm{c}}$ (shown in the figure), the threshold distance for contact formation. The dots are simulation results. The agreement between simulations and theory is excellent. Asymptotically $\langle R \rangle$ approaches $r_{\mathrm{c}}P^{-1/3}$ (dashed lines). The threshold for contact is expressed in terms of $a$ which is the equilibrium bond length in Equation \ref{eq:15}. (\textbf{c}) Illustration of the sensitivity of $r_{\mathrm{c}}$ in determining the mean spatial distance $\langle R\rangle$. Blue and yellow curves are computed by solving $\langle R\rangle$ (Equation (\ref{eq:1})) for a given contact probability $P_{mn}=10^{-3}$, and $r_{\mathrm{c}}$. The calculated $\langle R_{mn}\rangle$ is used in Equation \ref{eq:10} to obtain the distribution of the spatial distance $P(R_{mn})$. Blue and yellow curves are for the same value of $P$ but different $r_{\mathrm{c}}$ values.}
\clearpage

\begin{figure}[htb]
\centering
\includegraphics[width=0.7\textwidth]{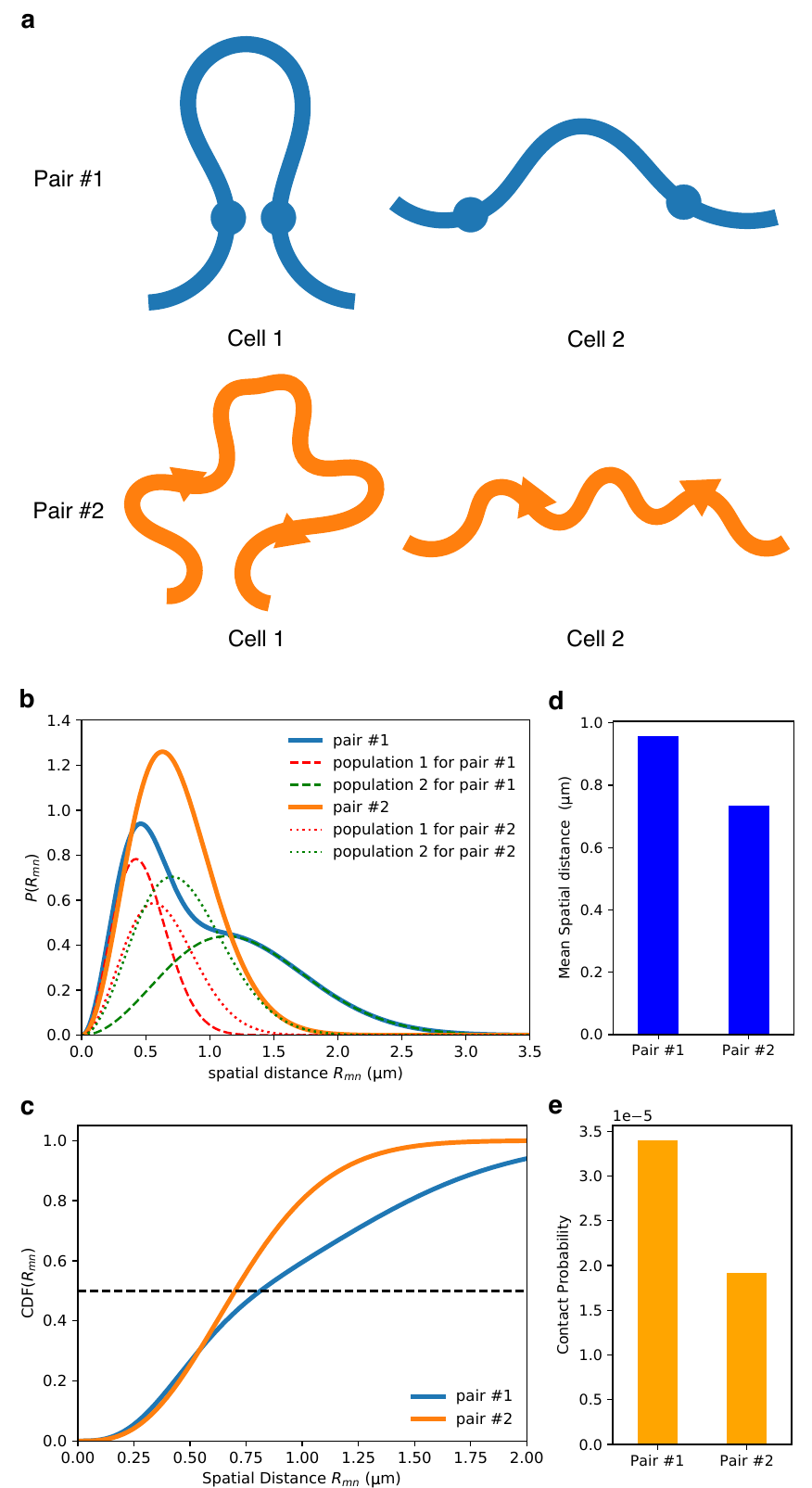}
\label{fig:fig2}
\end{figure}
\clearpage
\captionof{figure}{Illustrating the FISH-Hi-C ($[P_{mn},\langle R_{mn} \rangle]$) paradox. \textbf{(a)} Schematic illustration of the populations of two cells. There are two pairs of loci, pair 1 and pair 2. Cells 1 and 2 belong to two distinct populations such that pair 1 and pair 2 have different distributions of distances in the two cells. Pair 1 is always in proximity (contact is formed) in cell 1, whereas it is spatially separated (mean distance $> r_{\mathrm{c}}$) in cell 2. Pair 2, on the other hand, has similar distributions of spatial distance in cells 1 and 2. Cell with two different populations gives rise to paradoxical behavior, which is illustrated by choosing $\eta_{1}=0.4$ and $\eta_{2}=1-\eta_{1}=0.6$. These are the probabilities for a cell belonging to population 1 and 2, respectively. The pair 1 has parameters $\sigma_{1}=0.3\mu m$ and $\sigma_{2}=0.8\mu m$. The pair 2 has parameters $\sigma_{1}=0.4\mu m$ and $\sigma_{2}=0.5\mu m$. See Equation (\ref{eq:2}) for the definition of $\sigma_{1}$ and $\sigma_{2}$. \textbf{(b)} The distribution of distance for pair 1 (thick blue) and pair 2 (thick orange), respectively. The distributions for the two different populations are shown separately for pair 1 (dashed lines) and pair 2 (dotted lines). \textbf{(c)} Cumulative distribution of the spatial distance. The horizontal dashed line indicates the median distance. \textbf{(d)} Mean distances for pair 1 is larger than for pair 2. \textbf{(e)} Pair 1 has larger contact probability than 2, which is paradoxical since the distance between the loci in pair 1 is larger than in 2. The threshold for determining contact is $r_{\mathrm{c}}=20\mathrm{\ nm}$.}
\clearpage

\begin{figure}[htb]
\centering
\includegraphics[width=\textwidth]{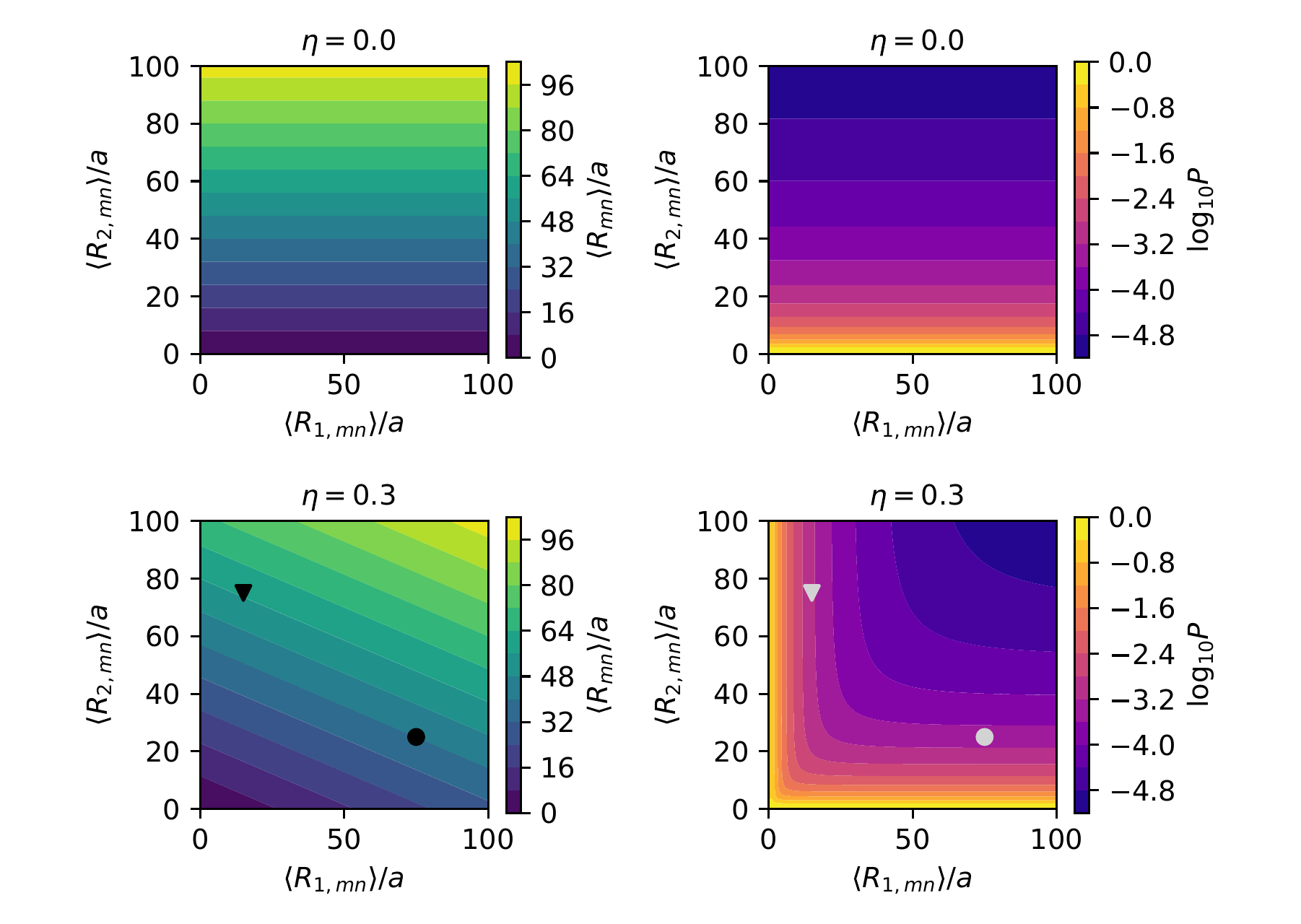}
\end{figure}
\captionof{figure}{Plots of mean distance $\langle R_{mn}\rangle$ and the contact probability $P_{mn}$ as heatmaps computed using $r_{\mathrm{c}}=2a$. The colorbars on the right show the values of $\langle R_{mn}\rangle$ and $P_{mn}$. The results for $\eta=0(\neq 0)$ is shown on top panel (bottom panel). Two specific pairs are marked as triangle and circle in the lower-left panel These loci pairs illustrate the $[P_{mn},\langle R_{mn} \rangle]$ paradox.}
\clearpage

\begin{figure}[htb]
\centering
\includegraphics[width=0.7\textwidth]{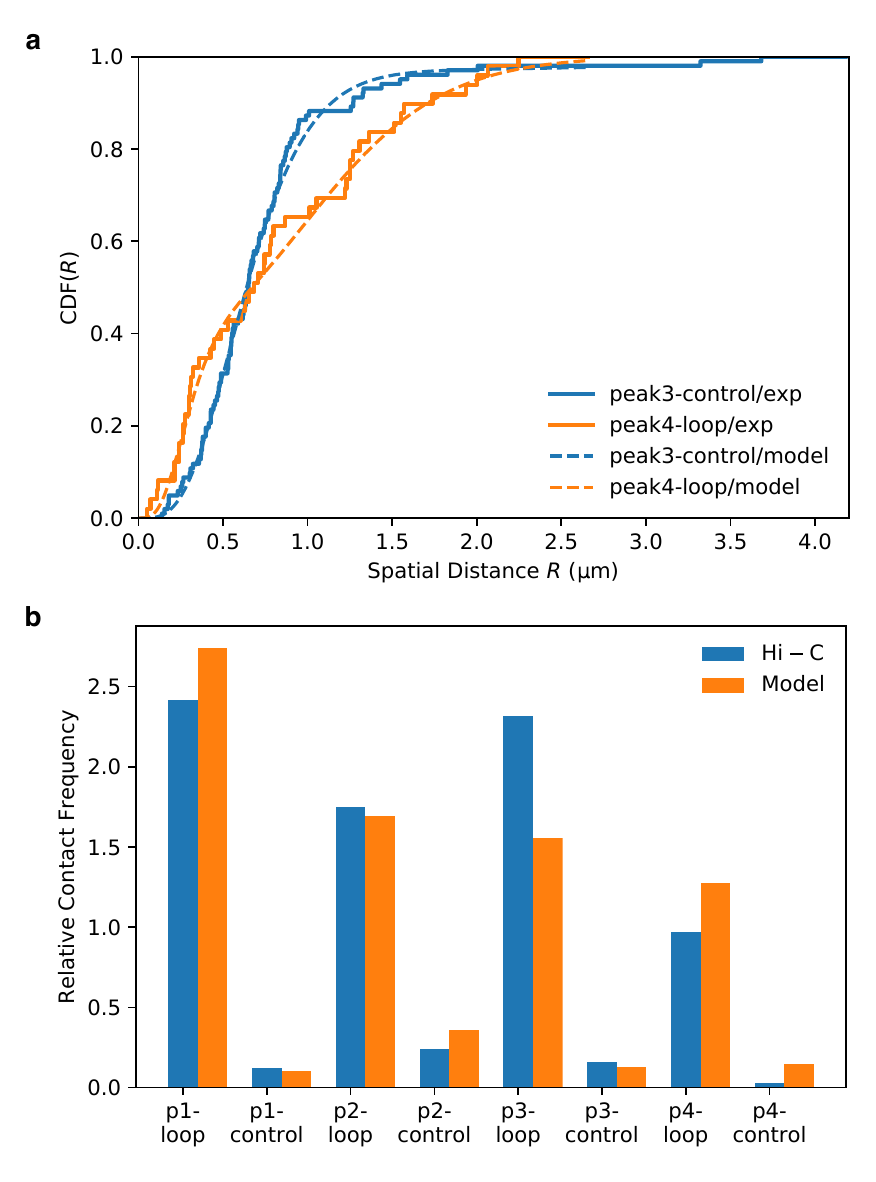}
\end{figure}
\captionof{figure}{Extracting statistics of subpopulations from FISH data. \textbf{(a)} Cumulative distribution function of the spatial distance, $\mathrm{CDF}(R)$ for two pairs of loci, labeled peak3-control and peak4-loop in \cite{rao20143d}. The excellent agreement between theory and experiments shows the usefulness of the relationship between $P_{mn}$ and $R_{mn}$ obtained using GRMC. The solid curves are the experiment data \cite{rao20143d}. The dashed lines are the fits to $\int_{0}^{R}P(r)\mathrm{d}r$ (the needed expressions are in Equation (\ref{eq:3}) and Supplementary Equation 1). The best fit parameters are: $\eta_{\mathrm{peak3-control}}\approx0.97$, $\langle R_{1,\mathrm{peak3-control}}\rangle \approx 0.67\mathrm{\ \mu m}$, $\langle R_{2,\mathrm{peak3-control}}\rangle \approx 4.08\mathrm{\ \mu m}$, $\eta_{\mathrm{peak4-loop}}\approx0.42$, $\langle R_{1,\mathrm{peak4-loop}}\rangle \approx 0.30\mathrm{\ \mu m}$ and $\langle R_{2,\mathrm{peak4-loop}}\rangle \approx 1.21\mathrm{\ \mu m}$. \textbf{(b)} Relative Contact Frequency computed from the fits of $\mathrm{CDF}(R)$ for eight pairs of loci investigate experimentally \cite{rao20143d} (orange bars). For each pair of loci, the contact probability is calculated as $P_{mn}=\int_{0}^{r_{\mathrm{c}}}P(r)\mathrm{d}r$ (Equation (\ref{eq:3})) using the parameters obtained by fitting $\mathrm{CDF}(R)$ with $r_{\mathrm{c}}=20\mathrm{\ nm}$. Comparison of the $\mathrm{CDF}(R)$s between theory and experiments for the eight pairs of loci are displayed in Supplementary Figure 2.  Blue bars are computed using the contact number from Hi-C measurements in \cite{rao20143d}. The relative contact frequency is calculated as $P_{i}/\langle P\rangle$ where $P_{i}$ is the contact probability computed using the model or the contact number measured in Hi-C for $i^{th}$ pair, and $\langle P\rangle$ is the mean value for all the pairs considered. p1-loop/p1-control/... are the ones referred to peak1-loop/peak1-control/... in \cite{rao20143d}.}
\clearpage

\begin{figure}[htb]
\centering
\includegraphics[width=\textwidth]{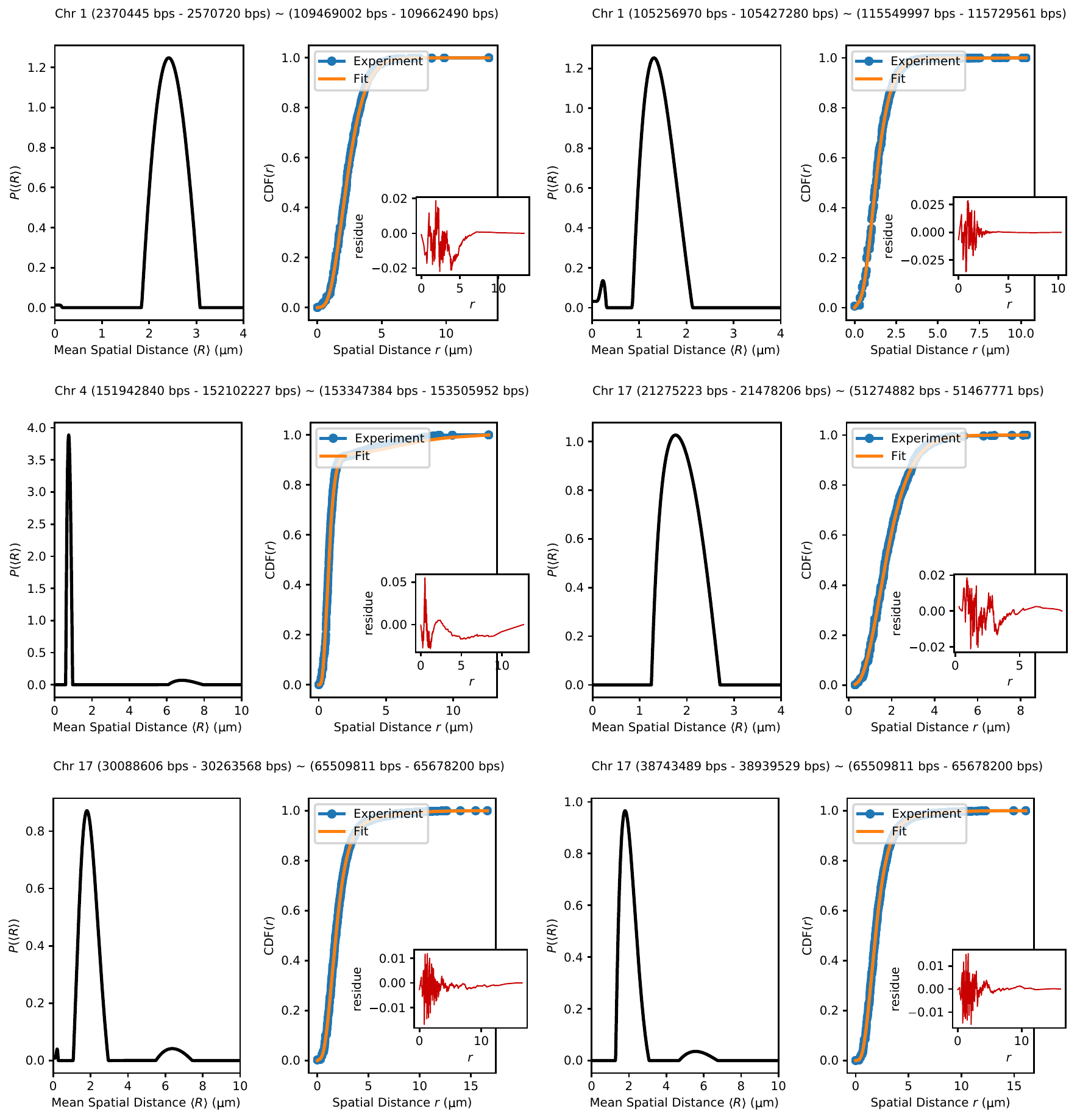}
\end{figure}
\captionof{figure}{Exampled fits of $\mathrm{CDF}(r)$ using Supplementary Equation 9 to the experimental data \cite{finn2019extensive}. The six exampled pairs of loci are indicated above each subfigure. Orange lines, showing the fits using our theory, is indistinguishable from the experiment (the differences between fitted and experimental curve are shown in the insets). The distribution $P(\langle R\rangle)$ given in the integral equation (Supplementary Equation 9) is solved using non-negative Tikhonov Regularization (Supplementary Note 7). As shown here, $P(\langle R\rangle)$ have multi-peaks and are widespread, which is a manifestation of heterogeneity. We set $g=1$ and $\delta=5/4$.}
\clearpage

\begin{figure}[h]
\centering
\includegraphics[width=\textwidth]{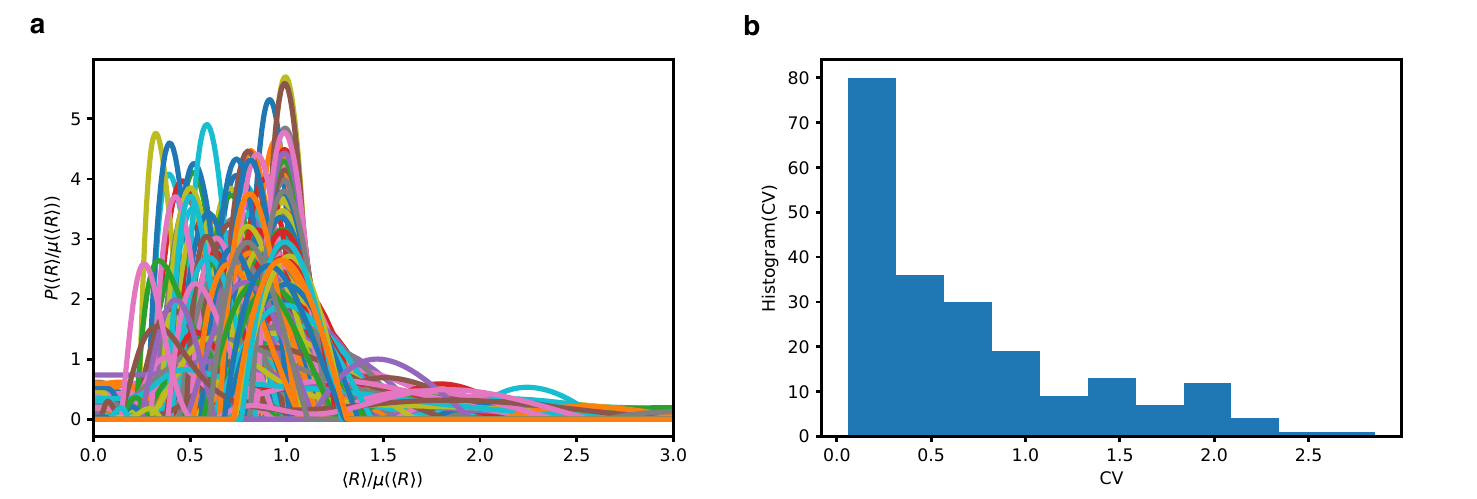}
\end{figure}
\captionof{figure}{Chromosome conformations are extensively heterogeneous. \textbf{(a)} Normalized distribution $P(\langle R\rangle /\mu(\langle R\rangle))$ ($\mu(\langle R\rangle)$ is the mean of $\langle R\rangle$) for all the 212 pairs of loci reported in \cite{finn2019extensive}.  For almost every pair of loci, the associated $P(\langle R\rangle /\mu(\langle R\rangle))$ has multiple peaks and is widespread. \textbf{(b)} Histogram of the coefficient of variations $\mathrm{CV}$ for all 212 pairs of loci probed in \cite{finn2019extensive}. The $\mathrm{CV}$ values are calculated for each pair of loci, using $\mathrm{CV} = \sigma(\langle R\rangle)/\mu (\langle R\rangle)$ where $\sigma(\langle R\rangle)$ is the standard deviation of $\langle R\rangle$. For a large number of loci pairs, CV exceeds 0.5, which is a quantitative measure of the  extensive heterogeneity noted in experiment \cite{finn2019extensive}}
\newpage

\begin{figure}[htb]
\centering
\includegraphics[width=0.7\textwidth]{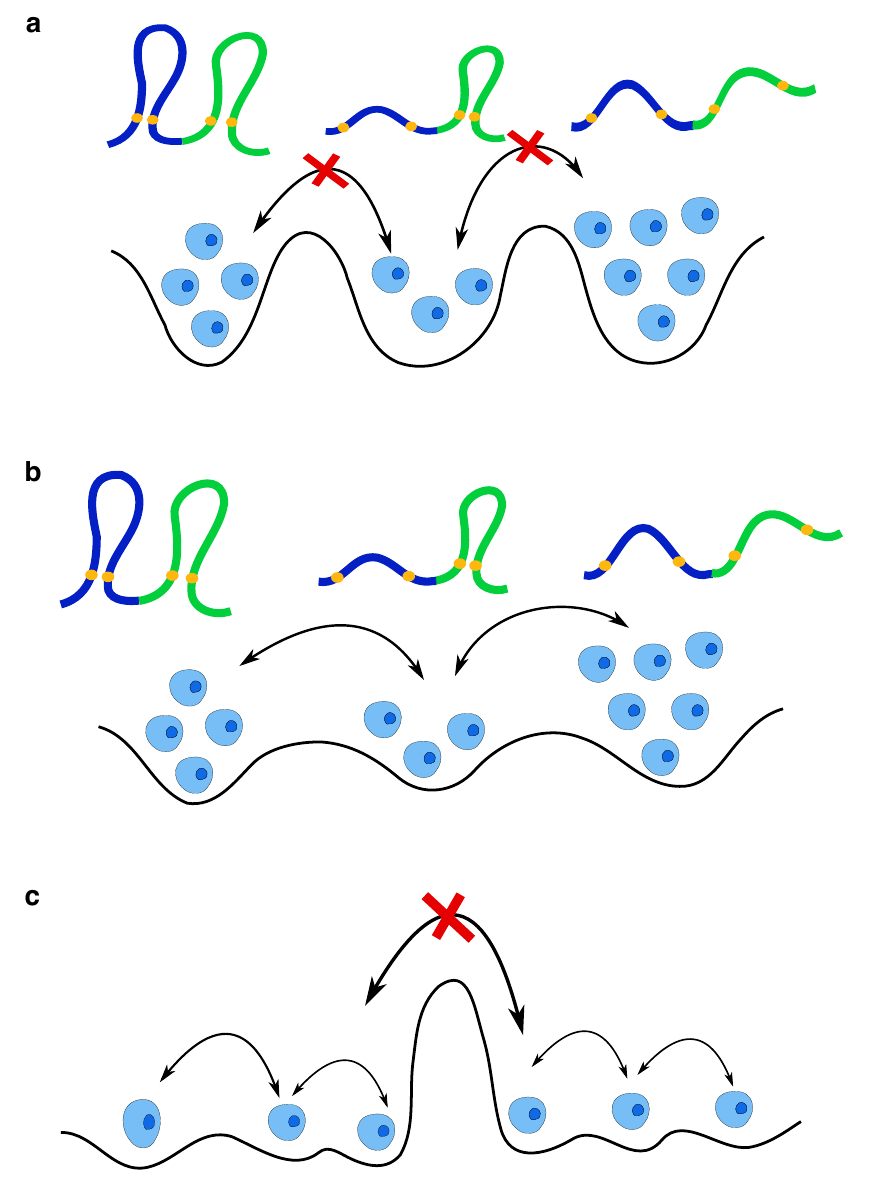}
\label{fig:fig5}
\end{figure}
\captionof{figure}{Schematic of the Genomic Folding Landscape (GFL). \textbf{(a)} Static heterogeneity: Cell subpopulation occupies distinct local minima in the GFL, with each minimum representing a stable organization. The energy barrier is too large for transitions between different local minima on a biological time scale (one cell cycle). \textbf{(b)} Dynamical heterogeneity: The energy barrier between local minima on the langscape is small enough which allows the dynamic transition between different subpopulations. \textbf{(c)} Combination of two different types of heterogeneity. In all three scenarios, the $[P_{mn},\langle R_{mn}\rangle]$ paradox arises. The loci contacts are in orange. The polymer conformation sketches are not shown in this scenario due to insufficient space.}
\clearpage

\end{document}